# Rectifying the phase-matching condition for fiber mode converter gratings using two-mode interference

**SOHAM BASU**

*École Polytechnique Fédérale de Lausanne, 1015 Lausanne*

*Corresponding author: sohambasu6817@gmail.com*

*Compiled August 13, 2020*

**The resonance wavelength, where a fiber mode converter grating written using periodic external perturbations achieves phase-matching, is both a critical design parameter and a device parameter. However, a method to precisely predict the resonance wavelength for any new fiber and grating writing apparatus has been missing so far. The missing link was the lack of direct experimental methods to estimate the modified intermodal phase after writing with external perturbations. The presented method can make this estimation from a single experiment, over a broad wavelength range, based on a novel mathematical connection between two-mode interference (TMI) and mode conversion. Using the novel methods, experimentally measured resonance wavelengths for different pitch and irradiation conditions have been predicted within relative errors of $4 \times 10^{-3}$.** © 2020 Optical Society of America

http://dx.doi.org/10.1364/ao.XX.XXXXXX

## 1. INTRODUCTION

Mode converter gratings (MC) in fibers had been widely researched and used, for sensing various modalities like temperature, strain, refractive index, bending, multi-parameter sensing etc [1], and for dispersion compensation in communication networks and laser cavities [2]. For waveguides along the axis $z$, MODES are finite number of orthogonal solutions to stationary Maxwell's equations at wavelength $\lambda$, with the following representation for the electric field

$$\begin{cases} \vec{E}_{\beta_k}(x,y,\lambda)e^{i(\omega t - \beta_k(\lambda)z)} & \text{propagating forward} \\ \vec{E}_{\beta_k}(x,y,\lambda)e^{i(\omega t + \beta_k(\lambda)z)} & \text{propagating backward} \end{cases} \quad (1)$$

where each mode can be identified by its unique propagation constant $\beta_k(\lambda) > 0$. The corresponding transverse electric field is given by $\vec{E}_{\beta_k}(x,y,\lambda) \in \mathbb{R}$. The symbol $t$ represents time, $\omega = \frac{2\pi c}{\lambda}$ represents frequency and $c$ is the speed of light in vacuum. For weakly-guiding fibers, such modes can be approximated by the LP approximation [3]. For example, for a step-index fiber the first 6 LP modes are abbreviated as $LP_{01}, LP_{11}, LP_{21}, LP_{02}, LP_{31}$ and $LP_{12}$, with the corresponding propagation constants satisfying $\beta_{01}(\lambda) < \beta_{11}(\lambda) < \beta_{21}(\lambda) < \beta_{02}(\lambda) < \beta_{31}(\lambda) < \beta_{12}(\lambda)$. $LP_{0l}$ modes have centrosymmetric mode profiles. Orthogonality of the modes implies that in absence of scattering, index perturbations or physical perturbations, they do not exchange power during propagation along the fiber. Even if the orthogonality is locally broken by local perturbations, the modes do not exchange significant power unless the effective wavevector of the periodic perturbation matches the difference in modified propagation constants of two modes at some wavelength (resonance wavelength $\lambda_{MC}$). Periodic local modifications change propagation constants locally, say to $\beta_1(\lambda) + \Delta\beta_1(\lambda, z, F)$ and $\beta_2(\lambda) + \Delta\beta_2(\lambda, z, F)$ for modes 1 and 2. Including the effect of such perturbations, the phase acquired by a mode with propagation constant $\beta_k(\lambda) + \Delta\beta_k(\lambda, z)$ over a period length $\Lambda$ is:

$$\phi_k(\lambda, F) = \int_z^{z+\Lambda} \left( \beta_k(\lambda) + \Delta\beta_k(\lambda, z', F) \right) dz', \text{ for } k = 1, 2 \quad (2)$$

where a period is starting at $z$, and $F = F(x, y, z')$, $z \leq z' \leq z + \Lambda$ signifies the refractive index perturbation in a period. A possible resonance wavelength (a) lies at the centre of mode conversion spectrum, and (b) satisfies the phase matching equation:

$$|\phi_1(\lambda_{MC}, F) - \phi_2(\lambda_{MC}, F)| = 2\pi \quad (3)$$

Writing the additional intermodal phase per period explicitly as $\Delta\phi(\lambda, F) = \int_z^{z+\Lambda} \left( \Delta\beta_1(\lambda, z', F) - \Delta\beta_2(\lambda, z', F) \right) dz'$, and the intermodal propagation constant difference for the pristine fiber as $\delta\beta_{1-2}(\lambda) = \beta_1(\lambda) - \beta_2(\lambda)$, the phase-matching condition from equation 3 can be rewritten in its rectified form:

$$|\Lambda \delta\beta_{1-2}(\lambda_{MC}) + \Delta\phi(\lambda_{MC}, F)| = 2\pi$$

$$\implies |\delta\beta_{1-2}(\lambda_{MC}) + \frac{\Delta\phi(\lambda_{MC}, F)}{\Lambda}| = \frac{2\pi}{\Lambda} \quad (4)$$

It has been long-understood that for MCs and long-period gratings (coupling forward-propagating core modes to cladding modes), $|\Lambda \delta\beta_{1-2}(\lambda_{MC})| \neq 2\pi$, or equivalently $\Delta\phi(\lambda_{MC}, F) \neq 0$. However, before this work, $\Delta\phi(\lambda, F)$ has never been experimentally accounted for due to lack of methods to measure it. Consequently, predicting the resonance wavelength was challenging, even when methods to estimate $\delta\beta_{1-2}(\lambda)$ have already been developed, for example by measuring two-mode interference (TMI) phase change during axial fiber stretching [4], or by using weak acousto-optic mode converter gratings of long length [5]. Even though detailed heuristic studies have been done on the dependence of the resonance wavelength on exposure conditions [6], an experimental algorithm which completely models such dependence was still to be discovered. The current work provides such a method, by combining equation 4 together with a novel method to measure $\Delta\phi(\lambda, F)$. Our novel method directly measures $\Delta\phi(\lambda, F)$ over the full wavelength range where TMI of the corresponding modes can be measured, by tracking unwrapped phase from TMI while writing such perturbations. As long as periodically written perturbations along the fiber axis are (a) non-overlapping, (b) identical in shape, and (c) having smaller perturbation footprint than the period length, $\Delta\phi(\lambda, F)$ only depends on the local perturbation and be independent of the period length. Such a perturbation is hereafter called a mark.

In section 2, we will show how to estimate $\Delta\phi(\lambda, F)$ by tracking unwrapped phase of TMI [7] while writing marks. Particularly, the technical prerequisites for reliable estimation will both be established and illustrated. Thereafter in section 3, we will show experimental verification of equation 4 with MCs fabricated using laser irradiation conditions with already characterized $\Delta\phi(\lambda, F)$, and different pitch. To strongly verify the method using equation 4, one of the MCs was fabricated to have resonance close to the turning-point wavelength [6], near which the resonance wavelength is highly sensitive to the pitch [8].

## 2. ESTIMATING PHASE ADDED PER MARK USING TMI

Two-mode interference is the characteristic oscillatory spectrum which is observed when two co-propagating fiber modes are simultaneously in-coupled and out-coupled, resulting in common-path Mach-Zhender interferometer [7]. The contrast is better when both the modes are excited with similar amplitudes. We achieve this by splicing a single-mode fiber which has comparable mode overlaps with the two TMI modes in the few-mode fiber (FMF) at the splice (Top panel of figure 1). Such a setup gives the following transmission intensity spectrum:

$$I_{TMI}(\lambda) = |A_{\beta_1}(\lambda)e^{i(\omega t - \beta_1(\lambda)z)} + A_{\beta_2}(\lambda)e^{i(\omega t - \beta_2(\lambda)z)}|^2$$
$$= |A_{\beta_1}(\lambda)|^2 + |A_{\beta_2}(\lambda)|^2 + 2\sqrt{A_{\beta_1}(\lambda)A_{\beta_2}(\lambda)}cos(\phi(\lambda))$$
$$= I_{dc}(\lambda) + I_{ac}(\lambda)cos(\phi(\lambda)) \quad \textbf{(5)}$$

where $\phi(\lambda) = \int_0^{L_{FMF}} \delta\beta_{1-2}(\lambda, z)\, dz$. $L_{FMF}$ is the few-mode fiber length. $\delta\beta_{1-2}(\lambda, z)$ accommodates perturbations, and equals $\delta\beta_{1-2}(\lambda)$ for the pristine fiber. $A_{\beta_1}(\lambda)$ and $A_{\beta_2}(\lambda)$ are products of mode overlaps at both splices and the input field amplitude:

$$A_{\beta_k}(\lambda) = \sqrt{I_{\beta_{SMF}}(\lambda)}\, \frac{<\vec{E}_{\beta_{SMF}}, \vec{E}_{\beta_k}>^2}{<\vec{E}_{\beta_{SMF}}, \vec{E}_{\beta_{SMF}}><\vec{E}_{\beta_k}, \vec{E}_{\beta_k}>},\, k=1,2$$
$$<\vec{E}_a, \vec{E}_b> = \int_{\mathbb{R}^2} \vec{E}_a(x, y, \lambda) \cdot \vec{E}_b(x, y, \lambda)\, dxdy \quad \textbf{(6)}$$

For individual spectra recorded after writing marks which do not satisfy the phase-matching condition, curve fitting algorithms can be run to fit equation 5 to a spectrum, with polynomial wavelength dependences for $I_{dc}(\lambda)$, $I_{ac}(\lambda)$ and $\phi(\lambda)$. For example, we performed this by using custom equation fitting in Matlab cftool. In order to initialize the fitting algorithm for each spectrum, extracted parameters from the previous spectrum can be as guess. For the first frame, $I_{dc}(\lambda)$ and $I_{ac}(\lambda)$ can be guessed by fitting the spectral extrema, which can be determined after fitting the noisy TMI signal with smoothing spline (for which we used Matlab cftool). For guessing $\phi(\lambda)$ of the first frame, polynomial fitting can be done after assigning phase of $2M\pi$ and $(2M+1)\pi$ to the maxima and minima wavelengths respectively, for appropriate $M \in \mathbb{Z}$. Once approximate polynomial fits are available for $I_{dc}(\lambda)$, $I_{ac}(\lambda)$ and $\phi(\lambda)$, those can be refined by custom fitting equation 5 to the first spectrum. The extracted phases for all the frames have the same indeterminacy of $2N\pi$ in the offset value of $\phi(\lambda)$, for some unknown integer $N$. Such indeterminacy is characteristic to phase-unwrapping of interferometric signals. If precise estimate of $L_{FMF}$ is available, it can be used to estimate the shape of $\delta\beta_{1-2}(\lambda)$ barring an UNKNOWN OFFSET by simply dividing the extracted phase $\phi(\lambda)$ by $L_{FMF}$.

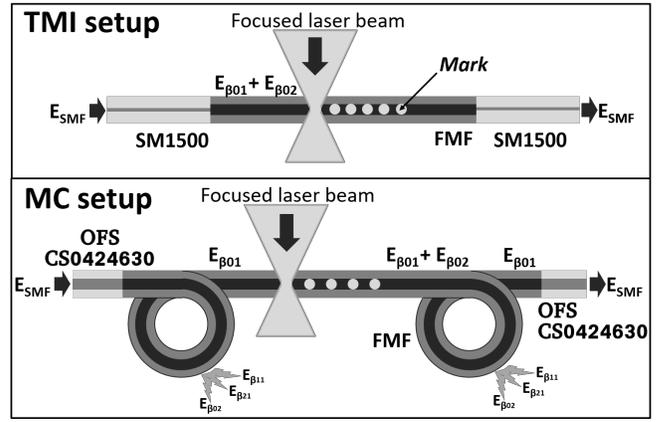

**Fig. 1.** TOP: Setup for measuring TMI spectral shift during writing non-resonant marks. BOTTOM: Setup for for measuring $LP_{01}$ transmission during mode converter grating writing. The SM1500 single-mode fiber has mode field diameter of 4.2 $\mu m$ at 1.55 $\mu m$. At centrosymmetric splices, this mode coupled to majorly $LP_{01}$ and $LP_{02}$ modes of the FMF (step-index research fiber; core diameter 10.0 $\mu m$, 13.7 % $GeO_2$ doping in the core) with approximately 70:30 ratio. The OFS CS0424630 single-mode fiber also coupled majorly to $LP_{01}$ and $LP_{02}$ modes of the FMF at centrosymmetric splices, but with ratio larger than 95:5. Making 20 loops of 7 mm diameter in the FMF got rid of all modes other than $LP_{01}$, below 40 dB extinction ratio.

For all experiments presented in this work, both TMI and MC, only $LP_{01}$ and $LP_{02}$ modes were excited in the FMF by centrosymmetric splicing. Keeping consistency with the equations presented beforehand, mode 1 is taken to be $LP_{01}$ and mode 2 is taken to be $LP_{02}$ such that $\beta_{01}(\lambda) > \beta_{02}(\lambda) \implies \delta\beta_{01-02} > 0$. In order to specifically differentiate between the estimate of the *shape* of $\delta\beta_{1-2}(\lambda)$ *with unknown offset* and the actual $\delta\beta_{1-2}(\lambda)$, we define a new quantity $\delta\beta^*_{01-02}(\lambda)$:

$$\delta\beta_{01-02}(\lambda) = \delta\beta^*_{01-02}(\lambda) + C \quad \textbf{(7)}$$

where $C$ is a constant which cannot be determined using TMI phase unwrapping, due to the offset indeterminacy of the phase. Although $\delta\beta^*(\lambda)$ could be estimated for each TMI sample which was irradiated, such estimate was prone to error since $L_{FMF}$ was small for those samples. Keeping $L_{FMF}$ smaller than the distance between the translation stage holder clamps (<4 cm) was necessary to avoid any changes in $\delta\beta(\lambda)$ due to bending of the FMF. To have more precise estimate of $\delta\beta^*_{01-02}(\lambda) = \frac{\phi(\lambda)}{L_{FMF}}$, a separate TMI sample with $L_{FMF} = 868 \pm 0.5$ mm was used [9].

Writing $N$ numbers of marks with the perturbation profile $F(x, y, z)$ adds extra phase $N\Delta\phi(\lambda, F)$ to the TMI phase, thereby causing shift in the spectral fringes. The bottom panel in figure 2 shows such measured shift in the concatenated spectra, recorded after writing each mark with condition $F_2$. The condition $F_2$ involved writing Gaussian-shaped marks along the fiber axis with $1/e^2$ diameter 59 $\mu$m, whereas a single mark was written by a 15 s exposure of a focused spot from Optec-LSV3 commercial excimer laser setup (248 nm wavelength). The laser spot intensity was $548 \pm 0.5$ Wcm$^{-2}$. The solid black curve in the top panel of figure 2 shows the initial TMI spectrum for no written marks, from which the phase unwrapping was started. The dashed black curve and solid gray curve show the spectrum after writing 50 and 100 marks, respectively.

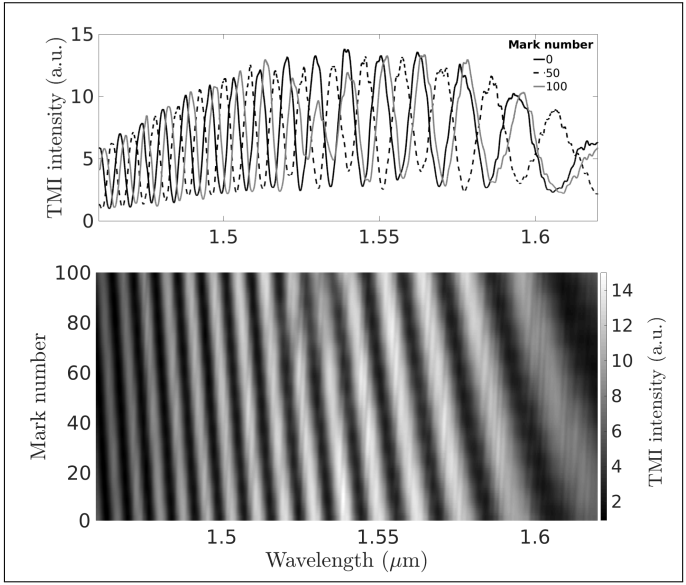

**Fig. 2.** TOP: TMI intensity spectrum for different mark numbers, for with condition $F_2$. BOTTOM: Shift of TMI fringe after writing each mark with condition $F_2$.

In practice the marks can have slight variability, but for repeatable mark writing the average phase added per mark should be the same for large number of marks. To estimate the average phase, the whole set of extracted phases for the first $N+1$ spectral frames is fitted with the following two-dimensional polynomial which is linear in the number of marks.

$$\hat{\phi}_N(\lambda, F) = \phi_{0,N}(\lambda, F) + (N-1)\Delta\phi_N(\lambda, F) \quad (8)$$

where $\phi_{0,N}(\lambda, F) = \sum_{m=0}^{d} c_{m,0}(N)\lambda^m$ and $\Delta\phi_N(\lambda, F) = \sum_{m=1}^{d} c_{m,1}(N)\lambda^{m-1}$ are polynomials only in wavelength. $d$ is the degree for the polynomial fitting of $\hat{\phi}(\lambda, 1)$ in terms of wavelength. The coefficients $c_{m,0}(N)$ and $c_{m,1}(N)$ might change with size of the spectral set, since these are outputs of the fitting algorithm. Such a fit for the unwrapped phase after writing $N$ marks provides an estimator for intermodal phase added per mark:

$$\Delta\phi_N(\lambda, F) = \frac{\hat{\phi}_N(\lambda, F) - \phi_{0,N}(\lambda, F)}{N-1} = \sum_{m=1}^{d} c_{m,1}(N)\lambda^{m-1} \quad (9)$$

Assuming repeatable marks (stability of the perturbation source parameters, negligible drift between the fiber and the perturbation source during mark writing), $\Delta\phi(\lambda, F)$ is the limit of the estimator in equation 9 with large number of marks

$$\Delta\phi(\lambda, F) = \lim_{N \to \infty} \Delta\phi_N(\lambda, F) \quad (10)$$

In order to verify convergence with number of written marks $N$, $\Delta\phi_N(\lambda, F)$ was plotted for fitting results for the spectral set corresponding to the first $N$ marks. The gray curves ($N = 10, 20, ..., 90$) in figure 3 illustrates such convergence of $\Delta\phi_N(\lambda, F_2)$ to the dashed black curve ($N = 100$) for the mark-writing condition $F_2$. Similar convergence was also checked for the mark writing condition $F_1$, which consisted of writing Gaussian-shaped marks along the fiber axis such that the $1/e^2$ diameter of a mark was roughly 25 $\mu$m, whereas a single mark was written by a 10 s exposure of a laser spot from the fourth-harmonic beam (257 nm wavelength, 180 fs pulsewidth, average power 40 mW) of Pharos femtosecond laser. The solid black curve in figure 3 illustrates the converged estimate $\Delta\phi_{200}(\lambda, F_1)$.

For an individual mark-writing condition, the number of marks $N$ was deemed sufficient for estimating $\Delta\phi_N(\lambda, F)$ when $|\Delta\phi_{N'}(\lambda, F) - \Delta\phi_{N'-1}(\lambda, F)| < 1 \times 10^{-4}$ radians was satisfied for $N' \geq N$. Then any $\Delta\phi_{N'}(\lambda, F)$ with $N' \geq N$ can be used as an estimate of $\Delta\phi(\lambda, F)$. Checking the convergence of $\Delta\phi_N(\lambda, F)$ is a prerequisite for reliable estimation of $\Delta\phi(\lambda, F)$.

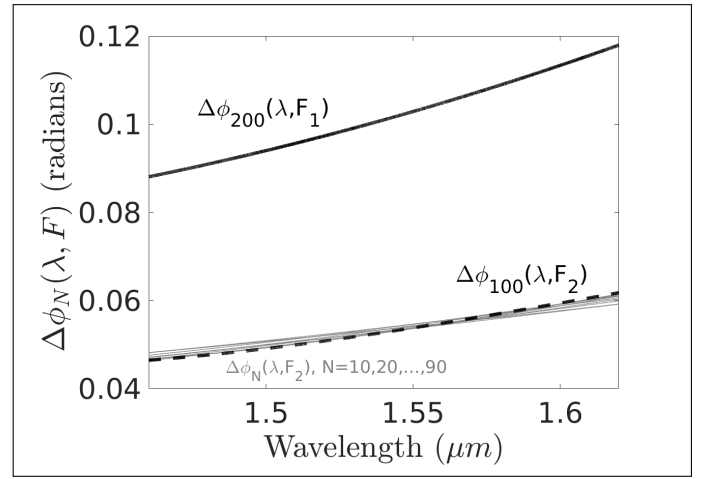

**Fig. 3.** Illustration of $\Delta\phi_{200}(\lambda, F_1)$ (solid black curve) and $\Delta\phi_{100}(\lambda, F_2)$ (dashed black curve). The gray curves highlight convergence of $\Delta\phi_N(\lambda, F_2)$ over intervals of 10 marks.

## 3. VERIFYING THE NOVEL PHASE-MATCHING EQUATION USING FABRICATED MCs

Once reliable estimate of $\Delta\phi(\lambda, F)$ is available for a certain mark-writing condition, equation 4 can be verified by fabricating MCs with the same condition. The setup for MC irradiation is illustrated in bottom panel of figure 1. Since $\delta\beta_{01-02}(\lambda) = \delta\beta^*_{01-02}(\lambda) + C > 0$, and $\Delta\phi(\lambda) > 0$ for all measurements, we rewrote equation 4 without the absolute sign:

$$\delta\beta_{01-02}(\lambda_{MC}) = \delta\beta^*_{01-02}(\lambda_{MC}) + C = \frac{2\pi - \Delta\phi(\lambda_{MC}, F)}{\Lambda} \quad (11)$$

where $\delta\beta^*_{01-02}(\lambda)$ and $\Delta\phi(\lambda), F)$ are known from TMI experiments, $\lambda_{MC}$ is measured from the MC spectrum during fabrication, and only the constant $C$ is unknown. We fixed $C$ such that it solved equation 11 at $\lambda_{MC} = \lambda_1$ for the irradiation condition $F_1$ and the pitch $\Lambda_1$ of the first fabricated MC:

$$\delta\beta^*_{01-02}(\lambda_1) + C = \frac{2\pi - \Delta\phi_{200}(\lambda_1, F_1)}{\Lambda_1}$$

$$\implies C = \frac{2\pi - \Delta\phi_{200}(\lambda_1, F_1)}{\Lambda_1} - \delta\beta^*_{01-02}(\lambda_1) \quad (12)$$

The intensity spectrum of $LP_{01}$ transmission corresponding to maximum mode conversion is shown as the solid black curve in figure 4. While writing marks with irradiation condition $F_1$ and pitch $\Lambda_1$, the resonance wavelength $\lambda_1$ was observed at $1.527 \pm 0.001\,\mu m$ for all marks.

Given $C$, the solid black curve in figure 5 illustrates $\delta\beta^*_{01-02}(\lambda) + C + \frac{\Delta\phi_{200}(\lambda, F_1)}{\Lambda_1}$ for irradiation condition $F_1$ and pitch $\Lambda_1$. The solid gray curve in figure 5 represents the determined $LP_{01}$-$LP_{02}$ propagation constant difference of the pristine fiber $\delta\beta(\lambda) = \delta\beta^*_{01-02}(\lambda) + C$.

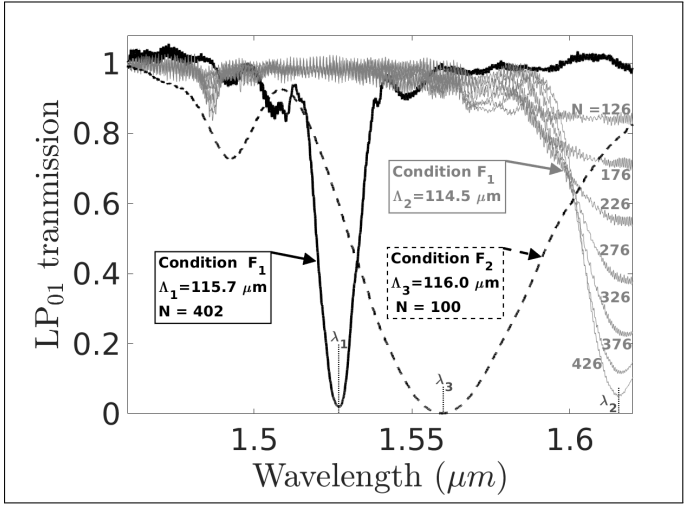

**Fig. 4.** Conversion spectra for different MCs written with mark-writing condition and pitch combinations $(F_1, \Lambda_1)$ (solid black curve), $(F_1, \Lambda_2)$ (solid gray curves) and $(F_2, \Lambda_3)$ (dashed black curve). For $\Lambda_2$, individual spectra for different number of marks (denoted by touching numbers) are shown to highlight that the resonance wavelength is not immediately distinguishable until the mode conversion spectrum is sufficiently deep near the turning-point wavelength. The spectral width for $F_2$ is broader due to its larger coupling constant [10], compared to $F_1$.

For another pitch $\Lambda_2$, at the same irradiation condition $F_1$, the phase matching curve would be $\delta\beta^*_{01-02}(\lambda) + C + \frac{\Delta\phi_{200}(\lambda, F_1)}{\Lambda_2}$. $\Lambda_2 = 114.5\,\mu m$ was chosen such that $\lambda_2$ was close to the $LP_{01}$-$LP_{02}$ turning-point wavelength at $1.639\,\mu m$ (where $\frac{\partial \delta\beta_{01-02}(\lambda)}{\partial \lambda} = 0$). The corresponding phase-matching curve is illustrated by the dashed gray curve in figure 5, which is almost indistinguishable from $\delta\beta^*_{01-02}(\lambda) + C + \frac{\Delta\phi_{200}(\lambda, F)}{\Lambda_1}$ as their difference $(\frac{1}{\Lambda_1} - \frac{1}{\Lambda_2})\Delta\phi_{200}(\lambda, F)$ is small in comparison. The corresponding resonance wavelength $\lambda_2$ is predicted to be at the root of the equation $\delta\beta^*_{01-02}(\lambda) + C + \frac{\Delta\phi_{200}(\lambda, F_1)}{\Lambda_2} = \frac{2\pi}{\Lambda_2}$, which has a value of $1.609\,\mu m$. The cross symbol in figure 5 illustrates the experimentally measured resonance wavelength $\lambda_2 = 1.615\,\mu m$ (solid gray curves in figure 4). The predicted value differed only by $0.006\,\mu m$, corresponding to a relative error $< 4 \times 10^{-3}$, even though being close to the turning-point wavelength [8].

For verification using another mark-writing condition $F_2$, $\Delta\phi_{100}(\lambda, F_2)$ was estimated using TMI, and thereafter another MC was fabricated using $F_2$ and a new pitch $\Lambda_3 = 116\,\mu m$. The measured resonance wavelength for $F_3$ and $\Lambda_3$ is measured to be $\lambda_3 = 1.560 \pm 0.002\,\mu m$ while writing the MC, which is illustrated by the "plus" symbol in figure 5. The calculated phase-matching curve for $F_3$ and $\Lambda_3$, given by $\delta\beta^*_{01-02}(\lambda) + C + \frac{\Delta\phi_{100}(\lambda, F_2)}{\Lambda_3}$ is illustrated by the dashed black curve in figure 5. This curve attains the value $\frac{2\pi}{\Lambda_3}$ at the predicted phase-matching wavelength of $1.564\,\mu m$. Also for the irradiation condition $F_2$, the relative error of the predicted resonance wavelength is only $< 3 \times 10^{-3}$.

For the presented irradiation conditions and pitch, using only $\delta\beta_{01-02}(\lambda) = \delta\beta^*_{01-02}(\lambda) + C = \frac{2\pi}{\Lambda}$ predicted no resonance in $1.46\,\mu m \leq \lambda \leq 1.62\,\mu m$. In contrast, the novel method predicted the MC resonance wavelengths with high accuracy.

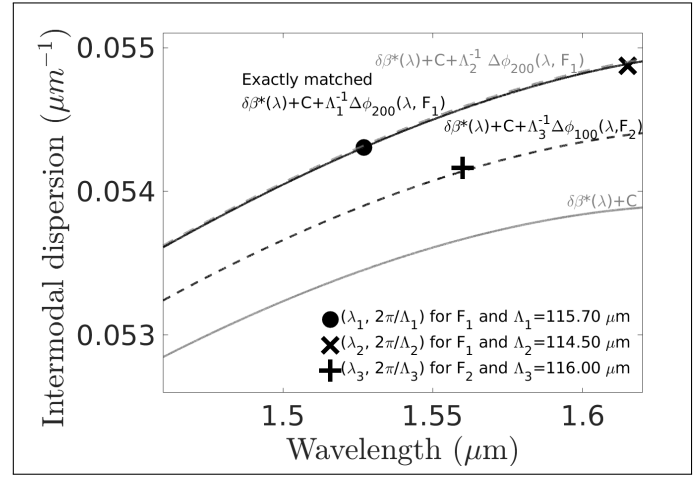

**Fig. 5.** Estimated phase-matching curves for different pitch ($\Lambda_1, \Lambda_2, \Lambda_3$) and mark-writing conditions ($F_1, F_2$), compared to fabricated MCs with these conditions.

## 4. CONCLUSION

We present a precise experimental method to predict the resonance wavelength for a mode converter grating written with any pitch, using any form of distinct spatially localized perturbations along the fiber axis. Predicted resonance wavelengths matched experimental measurements with relative errors $< 4 \times 10^{-3}$.

Using this method, the resonance wavelength can be predicted in a single experimental iteration, for any new fiber and mark-writing condition. This makes fabrication iterations unnecessary to achieve the desired resonance wavelength.

In addition, offset value of the propagation constant difference for the pristine fiber can also be accurately determined.

**Funding Information** Swiss National Science Foundation grant 200020_169415 is acknowledged for funding this project.